\documentclass[11pt]{article}
\usepackage{amsmath,amsfonts,amssymb,graphicx,epsfig}
\usepackage[usenames]{color}
\usepackage{pstricks}
\numberwithin{equation}{section}

\newcommand{\nc}{\newcommand}
\nc{\fh}{\hat{f}}
\nc{\muh}{\hat{\mu}}
\nc{\nuh}{\hat{\nu}}
\nc{\bib}{\bibitem}
\nc{\al}{\alpha}
\nc{\g}{\gamma}
\nc{\G}{\Gamma}
\nc{\D}{\Delta}
\nc{\eps}{\epsilon}
\nc{\la}{\lambda}
\nc{\La}{\Lambda}
\nc{\var}{\varphi}
\nc{\pa}{\partial}
\nc{\nn}{\nonumber \\ }
\nc{\hf}{\frac{1}{2}}
\nc{\dz}{\frac{dz}{2\pi i}}
\nc{\bin}[2]{\left(\!\!\!\begin{array}{c} {#1}\\ {#2} \end{array}\!\!\!\right)}
\nc{\be}{\begin{equation}}
\nc{\ee}{\end{equation}}
\nc{\bea}{\begin{eqnarray}}
\nc{\eea}{\end{eqnarray}}
\nc{\bra}[1]{\langle {#1}|}
\nc{\ket}[1]{|{#1}\rangle}
\nc{\chit}{\raisebox{0.25ex}{$\chi$}}
\nc{\Db}{\mbox{\boldmath $D$}}
\nc{\Hb}{\mbox{\boldmath $H$}}
\nc{\Hc}{{\cal H}}
\nc{\Rc}{{\cal R}}
\nc{\Lc}{{\cal L}}
\nc{\Vc}{{\cal V}}
\nc{\Ib}{\mbox{\boldmath $I$}}
\nc{\qb}{\bar{q}}
\def\vvdots{\mathinner{\mkern1mu\raise1pt\vbox{\kern7pt\hbox{.}}\mkern2mu
  \raise4pt\hbox{.}\mkern2mu\raise7pt\hbox{.}\mkern1mu}}
\def \st#1{\raisebox{-6pt}{\rule{0pt}{18pt}}\makebox[16pt]{\small ${#1}$}}
\nc{\gauss}[2]{\left[\!\!\begin{array}{c} {#1}\\ {#2} \end{array}\!\!\right]}
\nc{\sbin}[2]{\left\{\!\!\!\begin{array}{c} {#1}\\ {#2} 
\end{array}\!\!\!\right\}}
\nc{\sbinlr}[2]{\Big\langle\!\!\begin{array}{c} {#1}\\ {#2} 
\end{array}\!\!\Big\rangle}
\nc{\bino}[2]{\left(\!\!\begin{array}{c} {#1}\\ {#2} \end{array}\!\!\right)}

\definecolor{lightblue}{rgb}{.61,.61,1}
\definecolor{midblue}{rgb}{.7,.7,1}
\definecolor{lightlightblue}{rgb}{.85,.85,1}
\definecolor{lightestblue}{rgb}{.96,.96,1}

\nc{\ch}{{\rm ch}}
\nc{\R}{{\cal R}}
\nc{\dkk}{\delta_{j,\{k,k'\}}^{(2)}}
\nc{\ddkk}{\delta_{j,\{k,k'\}}^{(4)}}
\nc{\dddkk}{\delta_{j,\{k,k'\}}^{(8)}}

\begin{document}

\topmargin -5mm
\oddsidemargin 5mm

\setcounter{page}{1}

\mbox{}\vspace{-16mm}
\thispagestyle{empty}

\begin{center}
{\huge {\bf Fusion Algebra of Critical Percolation}}

\vspace{7mm}
{\Large J{\o}rgen Rasmussen}\ \ and\ \ {\Large Paul A. Pearce}\\[.3cm]
{\em Department of Mathematics and Statistics, University of Melbourne}\\
{\em Parkville, Victoria 3010, Australia}\\[.4cm]
J.Rasmussen@ms.unimelb.edu.au,\quad P.Pearce@ms.unimelb.edu.au

\end{center}

\vspace{8mm}
\centerline{{\bf{Abstract}}}
\vskip.4cm
\noindent
{\small We present an explicit conjecture for the chiral fusion algebra of critical percolation considering 
Virasoro representations with no enlarged or extended symmetry algebra. 
The representations we take to generate fusion are countably infinite in number. 
The ensuing fusion rules are quasi-rational in the sense that the fusion of a finite number of 
these representations decomposes into a finite direct sum of these representations. 
The fusion rules are 
commutative, associative and exhibit an $s\ell(2)$ structure.
They involve representations which we call Kac representations of which some are reducible yet indecomposable representations of rank 1. In particular, the identity of the fusion algebra
is a reducible yet indecomposable Kac representation of rank 1.
We make detailed comparisons of our fusion rules 
with the recent results of Eberle-Flohr and Read-Saleur. Notably, in agreement with Eberle-Flohr, 
we find the appearance of indecomposable representations of rank 3. 
Our fusion rules are supported by extensive numerical studies of an 
integrable lattice model of critical percolation. Details of our lattice findings and numerical results
will be presented elsewhere.}

\renewcommand{\thefootnote}{\arabic{footnote}}
\setcounter{footnote}{0}

\section{Introduction}

Percolation~\cite{Kesten82,Grimmet89,Stauffer92} has its origins in the paper~\cite{BroadHamm57} by Broadbent and Hammersley from 1957. Despite its relatively simple description, the subtleties and richness of percolation continue to hold much interest and even surprises after 50 years. One exciting recent development is the demonstration~\cite{Cardy92,Smirnov01} that the continuum scaling limit of percolation on the lattice yields a conformally invariant measure in the plane with connections to stochastic Loewner evolution~\cite{Schramm00,LawlerEtAl01,Werner03,RohdeEtAl05,KN,BB}. 
This is achieved by considering discrete analytic functions on the lattice. 
Another intriguing development is the unexpected connection~\cite{RazStrog01,BatchelorEtAl01,PearceEtAl02,FZZ06} between the groundstate of percolation, viewed as a stochastic process, and fully packed loop refinements of enumerations of symmetry classes of alternating sign matrices.

Percolation as a Conformal Field Theory (CFT) has some novel aspects being a non-rational and non-unitary theory with a countably infinite number of scaling fields. Most importantly, as argued 
in~\cite{Cardy99,GL,FL} for example, it is a {\em logarithmic} CFT with the consequence 
that it admits indecomposable representations of the Virasoro algebra~\cite{Roh96}. 
The first systematic study of logarithmic CFT appeared in~\cite{Gurarie93}. Logarithmic CFTs are currently the subject of intensive investigation, see~\cite{MS,Flohr97,RAK,Kausch00,bdylcft,MRS,Flohr03,Gaberdiel03,FFHST,Ruelle02,Kawai03,LMRS,Nichols02,RasWZW,FG,FGST,GR,QS} and references therein. 
There is of course a long history of studying percolation as the continuum scaling limit of lattice models~\cite{SaleurD87,SaleurSUSY,ReSa01}.
Here, however, it is convenient to regard critical percolation as a member of the family ${\cal LM}(p,p')$ of logarithmic CFTs defined as the continuum scaling limit of integrable lattice models~\cite{PRZ}. The first two members ${\cal LM}(1,2)$ and ${\cal LM}(2,3)$ correspond to critical dense polymers and critical percolation (bond percolation on the square lattice), 
respectively. This solvable model of critical dense polymers was considered in \cite{PRpoly}. 

In this paper, we are interested in the fusion algebra of ${\cal LM}(2,3)$ and we present
an explicit conjecture for the fusion rules generated from two fundamental representations, 
here denoted $(2,1)$ and $(1,2)$. The identity of this fundamental fusion algebra is denoted $(1,1)$
and is a reducible yet indecomposable representation of rank 1.
Our fusion rules are supported by extensive numerical studies of our 
integrable lattice model of critical percolation. Details of our lattice findings and numerical results
will be presented elsewhere.

It appears natural to suspect that the so-called augmented 
$c_{p,p'}$ models \cite{EberleF06} are equivalent to our logarithmic minimal models ${\cal LM}(p,p')$.
In particular, we believe that the augmented $c_{2,3}$ model is equivalent to critical percolation
${\cal LM}(2,3)$. 
Much is known~\cite{GK} about the fusion algebras of the 
augmented $c_{p,p'}$ models with $p=1$ while much less is known about the fusion algebras of these models for $p>1$. For critical percolation, the most complete information on fusion comes from Eberle and Flohr~\cite{EberleF06} who systematically applied the Nahm algorithm~\cite{Nahm94,GK} to obtain fusions level-by-level. 
A careful comparison shows that our fusion rules are compatible with their results~\cite{EberleF06}.
In particular, we confirm their observation of indecomposable representations of rank 3.
We also make a detailed comparison of our fusion rules with the results of \cite{RS} which we
find correspond to a subalgebra of our fusion algebra of critical percolation.

\subsection{Kac Representations}

Critical percolation ${\cal LM}(2,3)$ has central charge
$c=0$ and conformal weights
\be
 \D_{r,s}\ =\ \frac{(3r-2s)^{2}-1}{24},\hspace{1.2cm}r,s\in\mathbb{N}
\label{D}
\ee
The set of distinct conformal weights is $\{\D_{k,1},\D_{k+1,2},\D_{k+1,3};\ k\in\mathbb{N}\}
=\{\D_{1,k+1},\D_{2,k+2},;\ k\in\mathbb{N}\}$.

\begin{figure}[h]
\begin{center}
\begin{pspicture}(0,0)(7,8)
\rput[bl](0.15,0){\color{lightestblue}{\rule{6.9cm}{7.25cm}}}
\rput[bl](1.12,0){\color{lightlightblue}{\rule{.985cm}{7.25cm}}}
\rput[bl](3.09,0){\color{lightlightblue}{\rule{.985cm}{7.25cm}}}
\rput[bl](5.06,0){\color{lightlightblue}{\rule{.985cm}{7.25cm}}}
\rput[bl](0.15,1.32){\color{lightlightblue}{\rule{6.87cm}{.63cm}}}
\rput[bl](0.15,3.26){\color{lightlightblue}{\rule{6.87cm}{.63cm}}}
\rput[bl](0.15,5.20){\color{lightlightblue}{\rule{6.87cm}{.63cm}}}
\rput[bl](0.15,0){\color{lightblue}{\rule{.98cm}{1.3cm}}}
\rput[bl](1.11,1.3){\color{midblue}{\rule{.98cm}{.65cm}}}
\rput[bl](1.11,3.24){\color{midblue}{\rule{.98cm}{.65cm}}}
\rput[bl](1.11,5.18){\color{midblue}{\rule{.98cm}{.65cm}}}
\rput[bl](3.08,1.3){\color{midblue}{\rule{.98cm}{.65cm}}}
\rput[bl](3.08,3.24){\color{midblue}{\rule{.98cm}{.65cm}}}
\rput[bl](3.08,5.18){\color{midblue}{\rule{.98cm}{.65cm}}}
\rput[bl](5.05,1.3){\color{midblue}{\rule{.98cm}{.65cm}}}
\rput[bl](5.05,3.24){\color{midblue}{\rule{.98cm}{.65cm}}}
\rput[bl](5.05,5.18){\color{midblue}{\rule{.98cm}{.65cm}}}
\pswedge[fillstyle=solid,fillcolor=red,linecolor=red](1.1,1.92){.2}{180}{270}
\pswedge[fillstyle=solid,fillcolor=red,linecolor=red](2.09,1.92){.2}{180}{270}
\pswedge[fillstyle=solid,fillcolor=red,linecolor=red](1.1,3.86){.2}{180}{270}
\pswedge[fillstyle=solid,fillcolor=red,linecolor=red](2.09,3.86){.2}{180}{270}
\pswedge[fillstyle=solid,fillcolor=red,linecolor=red](1.1,5.80){.2}{180}{270}
\pswedge[fillstyle=solid,fillcolor=red,linecolor=red](2.09,5.80){.2}{180}{270}
\pswedge[fillstyle=solid,fillcolor=red,linecolor=red](2.09,.64){.2}{180}{270}
\pswedge[fillstyle=solid,fillcolor=red,linecolor=red](2.09,1.28){.2}{180}{270}
\pswedge[fillstyle=solid,fillcolor=red,linecolor=red](4.06,.64){.2}{180}{270}
\pswedge[fillstyle=solid,fillcolor=red,linecolor=red](4.06,1.28){.2}{180}{270}
\pswedge[fillstyle=solid,fillcolor=red,linecolor=red](4.06,1.92){.2}{180}{270}
\pswedge[fillstyle=solid,fillcolor=red,linecolor=red](6.03,.64){.2}{180}{270}
\pswedge[fillstyle=solid,fillcolor=red,linecolor=red](6.03,1.28){.2}{180}{270}
\pswedge[fillstyle=solid,fillcolor=red,linecolor=red](6.03,1.92){.2}{180}{270}
\rput[bl](0,0){
\begin{tabular}{|c|c|c|c|c|c|c|c|c|}
\hline
\st{\vdots}&\st{\vdots}&\st{\vdots}&\st{\vdots}&\st{\vdots}&\st{\vdots}&\st{\vvdots}\\ \hline
\st{12}&\st{\frac{65}{8}}&\st{5}&\st{\frac{21}{8}}&\st{1}&\st{\frac{1}{8}}&\st{\cdots}\\ \hline
\st{\frac{28}{3}}&\st{\frac{143}{24}}&\st{\frac{10}{3}}&\st{\frac{35}{24}}&\st{\frac{1}{3}}
&\st{-\frac{1}{24}}&\st{\cdots}\\ \hline
\st{7}&\st{\frac{33}{8}}&\st{2}&\st{\frac{5}{8}}&\st{0}&\st{\frac{1}{8}}&\st{\cdots}\\ \hline
\st{5}&\st{\frac{21}{8}}&\st{1}&\st{\frac{1}{8}}&\st{0}&\st{\frac{5}{8}}&\st{\cdots}\\ \hline
\st{\frac{10}{3}}&\st{\frac{35}{24}}&\st{\frac{1}{3}}
 &\st{-\frac{1}{24}}&\st{\frac{1}{3}}&\st{\frac{35}{24}}&\st{\cdots}\\ \hline
\st{2}&\st{\frac{5}{8}}&\st{0}&\st{\frac{1}{8}}&\st{1}&\st{\frac{21}{8}}&\st{\cdots}\\ \hline
\st{1}&\st{\frac{1}{8}}&\st{0}&\st{\frac{5}{8}}&\st{2}&\st{\frac{33}{8}}&\st{\cdots}\\ \hline
\st{\frac{1}{3}}&\st{-\frac{1}{24}}&\st{\frac{1}{3}}&\st{\frac{35}{24}}&\st{\frac{10}{3}}&\st{\frac{143}{24}}
&\st{\cdots}\\ \hline
\st{0}&\st{\frac{1}{8}}&\st{1}&\st{\frac{21}{8}}&\st{5}&\st{\frac{65}{8}}&\st{\cdots}\\ \hline
\st{0}&\st{\frac{5}{8}}&\st{2}&\st{\frac{33}{8}}&\st{7}&\st{\frac{85}{8}}&\st{\cdots}\\ \hline
\end{tabular}}
\end{pspicture}
\end{center}
\caption{Extended Kac table of critical percolation ${\cal LM}(2,3)$ showing the conformal weights $\Delta_{r,s}$ of the Kac  representations $(r,s)$. 
Except for the identifications $(2k,3k')=(2k',3k)$, the entries relate to 
{\em distinct} Kac representations even if the conformal weights coincide. This is unlike the
irreducible representations which are uniquely characterized by their conformal weight.  The periodicity of conformal weights $\Delta_{r,s}=\Delta_{r+2,s+3}$ is made manifest by shading the rows and columns with $r\equiv0$ (mod 2) or $s\equiv0$ (mod 3).  The Kac representations which happen to be irreducible representations are marked with a red shaded quadrant in the top-right corner. These do not exhaust the distinct values of the conformal weights. For example, the irreducible representation with $\Delta_{1,1}=0$ does not arise as a Kac representation. By contrast, the Kac table of the associated {\em rational} (minimal) model consisting of the shaded $1\times 2$ grid in the lower-left corner is trivial and contains only the operator corresponding to the irreducible representation with $\D=0$.}
\label{KacTable}
\end{figure}

{}From the lattice, a {\em Kac representation} $(r,s)$ arises for {\em every} pair of
integer Kac labels $r,s$ in the first quadrant
of the infinitely extended Kac table, see Figure~\ref{KacTable}. This relaxes the constraint $r=1,2$ considered in
\cite{PRZ}. The lattice description of the full set of Kac representations
will be discussed in detail elsewhere.
The conformal character of the Kac representation $(r,s)$ is given by
\be
 \chit_{r,s}(q)\ =\ \frac{q^{\frac{1}{24}+\D_{r,s}}}{\eta(q)}\left(1-q^{rs}\right)
\label{chikac}
\ee
where the Dedekind eta function is defined by
\be
 \eta(q)\ =\ q^{1/24}\prod_{m=1}^\infty(1-q^m)
\label{eta}
\ee
We will denote
the character of the {\em irreducible} Virasoro representation of conformal
weight $\D_{r,s}$ by $\ch_{r,s}(q)$. These irreducible characters \cite{FSZ} read
\bea
 {\rm ch}_{2k-1,a}(q)\!\!&=&\!\!K_{12,6k-3-2a;k}(q)-K_{12,6k-3+2a;k}(q),\hspace{1.2cm}a=1,2\nn
 {\rm ch}_{2k+1,3}(q)\!\!&=&\!\!
  \frac{1}{\eta(q)}\big(q^{3(2k-1)^2/8}-q^{3(2k+1)^2/8}\big)\nn
 {\rm ch}_{2k,b}(q)\!\!&=&\!\!
  \frac{1}{\eta(q)}\big(q^{(3k-b)^2/6}-q^{(3k+b)^2/6}\big),\hspace{2.1cm} b=1,2,3
\label{laq}
\eea
where $k\in\mathbb{N}$ while $K_{n,\nu;k}(q)$ is defined as
\be
 K_{n,\nu;k}(q)\ =\ \frac{1}{\eta(q)}\sum_{j\in\mathbb{Z}\setminus\{1,\ldots,k-1\}}q^{(nj-\nu)^2/2n}
\label{Kk}
\ee
It follows that for $k=1$, the first expression in (\ref{laq}) reduces to the well-known 
irreducible character 
\be
  {\rm ch}_{1,a}(q)\ =\ \frac{1}{\eta(q)}\sum_{j\in\mathbb{Z}}\big(q^{(12j-1)^2/24}-q^{(12j+5)^2/24}\big)
   \ =\ 1,
  \hspace{1.5cm} a=1,2
\label{r0s0}
\ee

A priori, a Kac representation can be either irreducible or reducible.
In the latter case, it could be fully reducible
(in which case it would be a direct sum of irreducible representations)
or its direct-sum
decomposition could involve at least one reducible but indecomposable representation
of rank 1 (possibly in addition to some irreducible representations).
We will only characterize the Kac representations appearing in the
fusion algebras to be discussed in the present work.
Among these are the Kac representations $\{(2k,1),(2k,2),(2k,3),(1,3k),(2,3k);\ k\in\mathbb{N}\}$.
Since their characters
all correspond to irreducible Virasoro characters, these Kac representations must
themselves be irreducible. They constitute an exhaustive list of irreducible Kac representations.
Two Kac representations are naturally identified if they have identical
conformal weights and are both irreducible. The relations
\be
 (2k,3)\ =\ (2,3k)
\label{idirr}
\ee
are the only such identifications. 
More general relations are considered in (\ref{kkexp}) and (\ref{RequalR}).
Here we merely point out that two Kac characters (\ref{chikac}) 
are equal $\chit_{r,s}(q)=\chit_{r',s'}(q)$ if and only if $(r',s')=(r,s)$ or $(r',s')=(2s/3,3r/2)$.
That is, the only equalities between Kac characters are of the form $\chit_{2k,3k'}(q)=\chit_{2k',3k}(q)$.
According to (\ref{RequalR}), a similar equality applies to the Kac representations themselves:
$(2k,3k')=(2k',3k)$.

The only {\em reducible} Kac representations entering the fundamental fusion algebra 
to be discussed below are $(1,1)$ and $(1,2)$
and they are both indecomposable representations of rank 1,
cf. Section \ref{sectionEF}.
The indecomposable representations of higher rank appearing in the fusion algebra
may be described in terms of Kac representations and their characters.
We therefore list the decompositions of the relevant Kac characters in terms of
irreducible characters
\bea
 \chit_{2k-1,b}(q)\!\!&=&\!\!
   \ch_{2k-1,b}(q)+\big(1-\delta_{b,3}\delta_{k,1}\big)\ch_{2k+1,b}(q),\hspace{3cm}b=1,2,3\nn
 \chit_{a,3k-b}(q)\!\!&=&\!\!\ch_{a,3k-b}(q)
   +\big(1-\delta_{a,2}\delta_{k,1}\big)\ch_{a,3k+b}(q),\hspace{3cm}a,b=1,2\nn
 \chit_{3,3k+b}(q)\!\!&=&\!\! \ch_{1,3k-3+b}(q)+\ch_{1,3k+b}(q)+\ch_{1,3k+3-b}(q)\nn
  \!\!&+&\!\!
   \ch_{1,3k+3+b}(q)+\ch_{1,3k+6-b}(q)+\ch_{1,3k+9-b}(q),\hspace{2cm}b=1,2
\label{chitch}
\eea
where $k\in\mathbb{N}$. The decomposition in the general case is discussed in the
appendix of \cite{PRZ}.

\section{Fusion Algebras}

The {\em fundamental} fusion algebra $\big\langle(2,1), (1,2)\big\rangle$ 
is defined as the fusion algebra
generated by the fundamental representations $(2,1)$ and $(1,2)$.
We find that closure of this fusion algebra requires the inclusion of a variety of
other representations
\be
  \big\langle(2,1), (1,2)\big\rangle\ =\ \big\langle(1,1), (1,2), (2k,a), (1,3k), (2k,3),
   \R_{2k,a}^{1,0}, \R_{2k,3}^{1,0}, \R_{a,3k}^{0,b},  
  \R_{2k,3}^{1,b};\ a,b=1,2;\ k\in\mathbb{N}\big\rangle
\label{A2112}
\ee
to be discussed next.

\subsection{Indecomposable Representations of Rank 2 or 3}

For $k\in\mathbb{N}$, the representations denoted by
$\R_{2k,1}^{1,0}$, $\R_{2k,2}^{1,0}$, $\R_{2k,3}^{1,0}$, $\R_{1,3k}^{0,1}$, 
$\R_{1,3k}^{0,2}$, $\R_{2,3k}^{0,1}$ and $\R_{2,3k}^{0,2}$ are indecomposable
representations of rank 2, while $\R_{2k,3}^{1,1}$ and $\R_{2k,3}^{1,2}$ are 
indecomposable representations of rank 3.
Their characters read
\bea
 \chit[\R_{2k,b}^{1,0}](q)\!\!&=&\!\!\chit_{2k-1,b}(q)+\chit_{2k+1,b}(q)\nn
  \!\!&=&\!\! \big(1-\delta_{b,3}\delta_{k,1}\big)\ch_{2k-1,b}(q)+2\ch_{2k+1,b}(q)
    +\ch_{2k+3,b}(q),\hspace{2cm}b=1,2,3\nn
 \chit[\R_{a,3k}^{0,b}](q)\!\!&=&\!\!\chit_{a,3k-b}(q)+\chit_{a,3k+b}(q)\nn
  \!\!&=&\!\!\big(1-\delta_{a,2}\delta_{k,1}\big)\ch_{a,3k-b}(q)+2\ch_{a,3k+b}(q)
    +\ch_{a,3(k+2)-b}(q),\hspace{1.2cm}a,b=1,2\nn
 \chit[\R_{2k,3}^{1,b}](q)\!\!&=&\!\!\chit_{2k-1,3-b}(q)+\chit_{2k-1,3+b}(q)
  +\chit_{2k+1,3-b}(q)+\chit_{2k+1,3+b}(q)\nn
  \!\!&=&\!\!  \big(1-\delta_{k,1}\big)\ch_{1,3k-3-b}(q)+2\big(1-\delta_{k,1}\big)\ch_{1,3k-3+b}(q)
   +2\ch_{1,3k-b}(q)\nn
 \!\!&+&\!\!4\ch_{1,3k+b}(q)+\big(2-\delta_{k,1}\big)\ch_{1,3k+3-b}(q)
  +2\ch_{1,3k+3+b}(q)\nn
 \!\!&+&\!\!2\ch_{1,3k+6-b}(q)+\ch_{1,3k+9-b}(q),\hspace{5.5cm}b=1,2
\label{chiR}
\eea
indicating that one may consider the indecomposable representations
as `indecomposable combinations' of Kac representations. The participating
Kac representations are of course the ones whose characters appear
in (\ref{chiR}). In the case of the indecomposable representation $\R_{2k,b}^{1,0}$
(or $\R_{a,3k}^{0,b}$) of rank 2, our lattice analysis indicates 
that a Jordan cell is formed between every state in $\ch_{2k+1,b}(q)$
(or $\ch_{a,3k+b}(q)$) and its partner state in the second copy of
$\ch_{2k+1,b}(q)$ (or $\ch_{a,3k+b}(q)$), and nowhere else.
In the case of the indecomposable representation $\R_{2k,3}^{1,b}$ of rank 3,
our lattice analysis indicates that for every quartet of matching states in the four 
copies of $\ch_{1,3k+b}(q)$, 
a rank-3 Jordan cell is formed along with a single state. It likewise
appears that a Jordan cell of rank 2 is formed between every pair of matching
states in the irreducible components with multiplicity 2.

The notation $\R_{r,s}^{a,b}$ is meant to reflect simple properties of the
higher-rank indecomposable representations. The pair of lower indices
thus refers to a `symmetry point' in the Kac table around which an indecomposable comination 
of Kac representations are located. The pair of upper indices indicates 
the distribution of these representations of which there are either two
(if $a=0$ or $b=0$) or four (if $a,b\neq0$). Their locations correspond to 
endpoints or corners, respectively, of a line segment or a rectangle with center
at $(r,s)$. This structure is encoded neatly in the character expressions (\ref{chiR}).

It follows from the lattice that the fundamental fusion algebra may be described
by separating the representations into a horizontal and a vertical part.
Before discussing implications of this, we examine the two directions individually, and
introduce some abbreviations.
To compactify the fusion rules, we use the notation
\be
 (r,-s)\ \equiv\ (-r,s)\ \equiv\ -(r,s),\hspace{1cm}\R_{-r,s}^{a,b}\ \equiv\ \R_{r,-s}^{a,b}\ 
   \equiv\ -\R_{r,s}^{a,b}
\ee
implying, in particular, that $(0,s)\equiv(r,0)\equiv\R_{0,s}^{a,b}\equiv\R_{r,0}^{a,b}\equiv0$,
and define the Kronecker delta combinations
\bea
 \dkk\!\!&=&\!\!  2-\delta_{j,|k-k'|}-\delta_{j,k+k'}  \nn
 \ddkk\!\!&=&\!\!   4-3\delta_{j,|k-k'|-1}-2\delta_{j,|k-k'|}-\delta_{j,|k-k'|+1}
   -\delta_{j,k+k'-1}-2\delta_{j,k+k'}-3\delta_{j,k+k'+1}\nn
 \dddkk\!\!&=&\!\!8-7\delta_{j,|k-k'|-2}-6\delta_{j,|k-k'|-1}-4\delta_{j,|k-k'|}-2\delta_{j,|k-k'|+1}
  -\delta_{j,|k-k'|+2}\nn
  \!\!&-&\!\!\delta_{j,k+k'-2}-2\delta_{j,k+k'-1}-4\delta_{j,k+k'}-6\delta_{j,k+k'+1}-7\delta_{j,k+k'+2}
\label{d24}
\eea

\subsection{Horizontal Fusion Algebra}

The {\em horizontal} fusion algebra $\big\langle(2,1)\big\rangle$ is defined as the fusion algebra
generated by the fundamental representation $(2,1)$. We find that closure 
of this fusion algebra requires the inclusion of the Kac representations $(2k,1)$
and the rank-2 indecomposable representations $\R_{2k,1}^{1,0}$ 
\be
 \big\langle(2,1)\big\rangle\ =\ \big\langle(2k,1), \R_{2k,1}^{1,0};\ k\in\mathbb{N}\big\rangle
\label{A21}
\ee
We conjecture that the fusion algebra $\big\langle(2,1)\big\rangle$ reads
\bea
 (2k,1)\otimes(2k',1)\!\!&=&\!\!\bigoplus_{j=|k-k'|+1,\ \!{\rm by}\ \!2}^{k+k'-1}
  \R_{2j,1}^{1,0}\nn
(2k,1)\otimes \R_{2k',1}^{1,0}\!\!&=&\!\!\bigoplus_{j=|k-k'|}^{k+k'}
  \dkk(2j,1)\nn
 \R_{2k,1}^{1,0}\otimes \R_{2k',1}^{1,0}\!\!&=&\!\!\bigoplus_{j=|k-k'|}^{k+k'}
  \dkk\R_{2j,1}^{1,0}
   \label{fusion21}
\eea
This fusion algebra does not contain an identity.

\subsection{Vertical Fusion Algebra}

The {\em vertical} fusion algebra $\big\langle(1,2)\big\rangle$ is defined as the fusion algebra
generated by the fundamental representation $(1,2)$. We find that closure 
of this fusion algebra requires the inclusion of the Kac representations $(1,1)$ and
$(1,3k)$ and the rank-2 indecomposable representations $\R_{2k,1}^{0,b}$ 
\be
 \big\langle(1,2)\big\rangle
  \ =\ \big\langle(1,1), (1,2), (1,3k), \R_{1,3k}^{0,b};\ b=1,2;\ k\in\mathbb{N}\big\rangle
\label{A12}
\ee
Letting $X$ denote any of these representations, we conjecture that the fusion algebra
$\big\langle(1,2)\big\rangle$ reads
\bea
 (1,1)\otimes X\!\!&=&\!\!X\nn
 (1,2)\otimes (1,2)\!\!&=&\!\!(1,1)\oplus(1,3)\nn
 (1,2)\otimes (1,3k)\!\!&=&\!\!\R_{1,3k}^{0,1}\nn
 (1,2)\otimes \R_{1,3k}^{0,1}\!\!&=&\!\!\R_{1,3k}^{0,2}\oplus2(1,3k)\nn
 (1,2)\otimes \R_{1,3k}^{0,2}\!\!&=&\!\!\R_{1,3k}^{0,1}\oplus(1,3(k-1))\oplus (1,3(k+1))\nn
 (1,3k)\otimes(1,3k')\!\!&=&\!\!
  \bigoplus_{j=|k-k'|+1,\ \!{\rm by}\ \!2}^{k+k'-1}\big(\R_{1,3j}^{0,2}\oplus(1,3j)\big)\nn
 (1,3k)\otimes \R_{1,3k'}^{0,1}\!\!&=&\!\!
   \Big(\bigoplus_{j=|k-k'|+1,\ \!{\rm by}\ \!2}^{k+k'-1}2\R_{1,3j}^{0,1}\Big)
  \oplus\Big(\bigoplus_{j=|k-k'|,\ \!{\rm by}\ \!2}^{k+k'}\dkk (1,3j)\Big)\nn
 (1,3k)\otimes \R_{1,3k'}^{0,2}\!\!&=&\!\!  
  \Big(\bigoplus_{j=|k-k'|,\ \!{\rm by}\ \!2}^{k+k'}\dkk\R_{1,3j}^{0,1}\Big)
  \oplus\Big(\bigoplus_{j=|k-k'|+1,\ \!{\rm by}\ \!2}^{k+k'-1}2(1,3j)\Big)\nn
 \R_{1,3k}^{0,1}\otimes \R_{1,3k'}^{0,1}\!\!&=&\!\!\ 
  \Big(\bigoplus_{j=|k-k'|,\ \!{\rm by}\ \!2}^{k+k'}\dkk\R_{1,3j}^{0,1}\Big)
  \oplus\Big(\bigoplus_{j=|k-k'|+1,\ \!{\rm by}\ \!2}^{k+k'-1}\big(2\R_{1,3j}^{0,2}\oplus4(1,3j)\big)\Big)\nn
 \R_{1,3k}^{0,1}\otimes \R_{1,3k'}^{0,2}\!\!&=&\!\!
  \Big(\bigoplus_{j=|k-k'|+1,\ \!{\rm by}\ \!2}^{k+k'-1}2\R_{1,3j}^{0,1}\Big)
  \oplus 
   \Big(\bigoplus_{j=|k-k'|,\ \!{\rm by}\ \!2}^{k+k'}\dkk\big(\R_{1,3j}^{0,2}
   \oplus2(1,3j)\big)\Big)\nn
 \R_{1,3k}^{0,2}\otimes R_{1,3k'}^{0,2}\!\!&=&\!\! 
 \Big(\bigoplus_{j=|k-k'|,\ \!{\rm by}\ \!2}^{k+k'}\dkk\R_{1,3j}^{0,1}\Big)
  \oplus\Big(\bigoplus_{j=|k-k'|+1,\ \!{\rm by}\ \!2}^{k+k'-1}2\R_{1,3j}^{0,2}\Big)\nn
\!\!&\oplus&\!\! \Big(\bigoplus_{j=|k-k'|-1,\ \!{\rm by}\ \!2}^{k+k'+1}
  \ddkk(1,3j)\Big)
\label{fusion12}
\eea
It is noted that for $j=|k-k'|-1$ (mod 2), as in $\R_{1,3k}^{0,2}\otimes R_{1,3k'}^{0,2}$, 
the fusion multiplicity $\ddkk$ reduces to 
$4-3\delta_{j,|k-k'|-1}-\delta_{j,|k-k'|+1}-\delta_{j,k+k'-1}-3\delta_{j,k+k'+1}$.
The representation $(1,1)$ is the identity of this vertical fusion algebra.

\subsection{Comparison with Read and Saleur}

It is verified that
\be
 \big\langle(1,1),(1,6k-3),\R_{1,6k}^{0,1},\R_{1,6k-3}^{0,2};\ k\in\mathbb{N}\big\rangle
\label{subvert}
\ee
is a subalgebra of the vertical fusion algebra. It corresponds to the fusion
algebra of critical percolation discussed by Read and Saleur in \cite{RS}. To appreciate this, we 
provide a dictionary for translating the representations generating the subalgebra
(\ref{subvert}) into the notation used in \cite{RS}
\bea
 (1,1)\ \ &\longleftrightarrow&\ \ \R_0\nn
 (1,2j+1)\ \ &\longleftrightarrow&\ \ \R_j,\hspace{1cm}j\equiv1\ ({\rm mod}\ 3)\nn
 \R_{1,2j-1}^{0,2}\ \ &\longleftrightarrow&\ \ \R_j,\hspace{1cm}j\equiv2\ ({\rm mod}\ 3)\nn
 \R_{1,2j}^{0,1}\ \ &\longleftrightarrow&\ \ \R_j,\hspace{1cm}j\equiv0\ ({\rm mod}\ 3)
\label{dictRS}
\eea
where $j\in\mathbb{N}$. We find that their fusion algebra is in agreement with 
the subalgebra (\ref{subvert}) of the vertical fusion algebra $\big\langle(1,2)\big\rangle$ which itself
is a subalgebra of the fundamental fusion algebra $\big\langle(2,1),(1,2)\big\rangle$ of critical
percolation.

\subsection{Fundamental Fusion Algebra}

It follows from the lattice description that the fundamental fusion algebra
$\big\langle(2,1),(1,2)\big\rangle$ is both associative and commutative.
As already announced, it also follows from the lattice that the representations may be separated
into a horizontal and a vertical part. For the Kac representations, this implies
\be
 (r,s)\ =\ (r,1)\otimes(1,s)
\label{r11s}
\ee
For the purposes of examining the fundamental fusion algebra, we introduce
the representations
\bea
 (2k,3k')\!\!&=&\!\!(2k,1)\otimes(1,3k'),\hspace{1.5cm}
 \R_{2k,3k'}^{1,0}\ =\ \R_{2k,1}^{1,0}\otimes(1,3k')\nn
 \R_{2k,3k'}^{0,b}\!\!&=&\!\!(2k,1)\otimes \R_{1,3k'}^{0,b},\hspace{1.67cm}
 \R_{2k,3k'}^{1,b}\ =\ \R_{2k,1}^{1,0}\otimes \R_{1,3k'}^{0,b}
\label{kk}
\eea
thus defined as the result of certain simple fusions of `a horizontal and a vertical representation'.
As we will show elsewhere, these representations may be decomposed 
in terms of the representations listed in (\ref{A2112})
\bea
 (2k,3k')\!\!&=&\!\!\bigoplus_{j=|k-k'|+1,\ \!{\rm by}\ \!2}^{k+k'-1}(2j,3),\hspace{1.5cm}
 \R_{2k,3k'}^{1,0}\ =\ \bigoplus_{j=|k-k'|+1,\ \!{\rm by}\ \!2}^{k+k'-1}\R_{2j,3}^{1,0}\nn
 \R_{2k,3k'}^{0,b}\!\!&=&\!\!\bigoplus_{j=|k-k'|+1,\ \!{\rm by}\ \!2}^{k+k'-1}\R_{2,3j}^{0,b},\hspace{1.55cm}
 \R_{2k,3k'}^{1,b}\ =\ \bigoplus_{j=|k-k'|+1,\ \!{\rm by}\ \!2}^{k+k'-1}\R_{2j,3}^{1,b}
\label{kkexp}
\eea
with
\be
 (2k,3k')\ =\ (2k',3k),\ \ \ \ \ \R_{2k,3k'}^{1,0}\ =\ \R_{2k',3k}^{1,0},\ \ \ \ \ 
   \R_{2k,3k'}^{0,b}\ =\ \R_{2k',3k}^{0,b},\ \ \ \ \ \R_{2k,3k'}^{1,b}\ =\ \R_{2k',3k}^{1,b}
\label{RequalR}
\ee 
as special identifications extending the set (\ref{idirr}). 
The fundamental fusion algebra is now obtained by simply
applying (\ref{kk}) and (\ref{kkexp}) to the fusion of a pair of representations in
(\ref{A2112}).
We illustrate this with a general but somewhat formal evaluation where we let 
$A_{r,s}=\bar{a}_{r,1}\otimes\ a_{1,s}$, $B_{r',s'}=\bar{b}_{r',1}\otimes\ b_{1,s'}$,
$\bar{a}_{r,1}\otimes\bar{b}_{r',1}=\bigoplus_{r''}\bar{c}_{r'',1}$ and
$a_{1,s}\otimes b_{1,s'}=\bigoplus_{s''}c_{1,s''}$. Our fusion prescription now yields
\bea
 A_{r,s}\otimes B_{r',s'}\!\!&=&\!\!\Big(\bar{a}_{r,1}\otimes a_{1,s}\Big)\otimes
  \Big(\bar{b}_{r',1}\otimes b_{1,s'}\Big)
   \ =\ \Big(\bar{a}_{r,1}\otimes\bar{b}_{r',1}\Big)\otimes
  \Big(a_{1,s}\otimes b_{1,s'}\Big)\nn
 \!\!&=&\!\!\Big(\bigoplus_{r''}\bar{c}_{r'',1}\Big)\otimes\Big(\bigoplus_{s''}c_{1,s''}\Big)
  \ =\ \bigoplus_{r'',s''}C_{r'',s''}
\label{rs}
\eea
where $C_{r'',s''}=\bar{c}_{r'',1}\otimes c_{1,s''}$.
Using this, the fundamental fusion algebra $\big\langle(2,1),(1,2)\big\rangle$ follows straightforwardly
from the fusion algebras $\big\langle(2,1)\big\rangle$ and $\big\langle(1,2)\big\rangle$ together with
(\ref{kk}) and (\ref{kkexp}). In particular, it follows readily that the Kac representation 
$(1,1)$ is the {\em identity} of the fundamental fusion algebra $\big\langle(2,1),(1,2)\big\rangle$.

In this brief communication, we will only apply this fusion prescription explicitly 
to the fusion of the two rank-2 indecomposable representations
$\R_{2k,2}^{1,0}$ and $\R_{2,3k'}^{0,2}$
\bea
 \R_{2k,2}^{1,0}\otimes\R_{2,3k'}^{0,2}\!\!&=&\!\!\Big(\R_{2k,1}^{1,0}\otimes (1,2)\Big)
   \otimes\Big((2,1)\otimes \R_{1,3k'}^{0,2}\Big)
   \ =\ \Big(\R_{2k,1}^{1,0}\otimes (2,1)\Big)\otimes\Big((1,2)\otimes \R_{1,3k'}^{0,2}\Big)\nn
 \!\!&=&\!\!\Big((2(k-1),1)\oplus2(2k,1)\oplus(2(k+1),1)\Big)\otimes\Big(\R_{1,3k'}^{0,1}\oplus
   (1,3(k'-1))\oplus(1,3(k'+1))\Big)\nn
 \!\!&=&\!\!\Big(\bigoplus_{j=|k-k'|}^{k+k'}\dkk\R_{2,3j}^{0,1}\Big)
    \oplus\Big(\bigoplus_{j=|k-k'|-1}^{k+k'+1}\delta_{j,\{k,k'\}}^{(4)}(2j,3)\Big)
\label{ex22}
\eea
and to the fusion of two rank-3 indecomposable representations
\bea
 \R_{2k,3}^{1,1}\otimes \R_{2k',3}^{1,1}\!\!&=&\!\!\Big(\R_{2k,1}^{1,0}\otimes \R_{1,3}^{0,1}\Big)
   \otimes\Big(\R_{2k',1}^{1,0}\otimes \R_{1,3}^{0,1}\Big)
   \ =\ \Big(\R_{2k,1}^{1,0}\otimes \R_{2k',1}^{1,0}\Big)
    \otimes\Big(\R_{1,3}^{0,1}\otimes \R_{1,3}^{0,1}\Big)\nn
 \!\!&=&\!\!\Big(\bigoplus_{j=|k-k'|}^{k+k'}\dkk\R_{2j,1}^{1,0}\Big)
   \otimes\Big(\R_{1,6}^{0,1}\oplus2\R_{1,3}^{0,2}\oplus4(1,3)\Big)\nn
 \!\!&=&\!\!\Big(\bigoplus_{j=|k-k'|-1}^{k+k'+1}\ddkk\R_{2j,3}^{1,1}\Big)
    \oplus\Big(\bigoplus_{j=|k-k'|}^{k+k'}\dkk\big(2\R_{2j,3}^{1,2}\oplus4\R_{2j,3}^{1,0}\big)\Big)
\label{ex3311}
\eea
and likewise
\bea
 \R_{2k,3}^{1,1}\otimes \R_{2k',3}^{1,2}\!\!&=&\!\! 
  \Big(\bigoplus_{j=|k-k'|}^{k+k'}\dkk2\R_{2j,3}^{1,1}\Big)
    \oplus\Big(\bigoplus_{j=|k-k'|-1}^{k+k'+1}\ddkk\big(\R_{2j,3}^{1,2}\oplus2\R_{2j,3}^{1,0}\big)\Big)  \nn
 \R_{2k,3}^{1,2}\otimes \R_{2k',3}^{1,2}\!\!&=&\!\!
  \Big(\bigoplus_{j=|k-k'|-1}^{k+k'+1}\ddkk\R_{2j,3}^{1,1}\Big)
  \oplus\Big(\bigoplus_{j=|k-k'|}^{k+k'}\dkk2\R_{2j,3}^{1,2}\Big)
  \oplus\Big(\bigoplus_{j=|k-k'|-2}^{k+k'+2}\dddkk\R_{2j,3}^{1,0}\Big)\nn
\label{ex3312}
\eea

Several subalgebras of the fundamental fusion algebra
are easily identified. An interesting example is the one generated by the set of
rank-3 indecomposable representations and the rank-2 indecomposable
representations $R_{2k,3}^{1,0}$.
Two other noteworthy subalgebras are the ones generated by all the representations
in (\ref{A2112}) except $(1,2)$ or $(1,1)$ and $(1,2)$.

We wish to point out that, at the level of Kac characters, the horizontal, vertical and
fundamental fusion algebras are all compatible with the $s\ell(2)$ structure
\be
 \phi_n\otimes\phi_{n'}\ =\ \bigoplus_{m=|n-n'|+1,\ \!{\rm by}\ \!2}^{n+n'-1}\phi_m
\label{sl2}
\ee
This is straightforward to establish for the horizontal and vertical fusion algebras
as illustrated by the fusion $\R_{2k,1}^{1,0}\otimes\R_{2k',1}^{1,0}$ where (\ref{sl2}) yields
\bea
 \chit[\R_{2k,1}^{1,0}\otimes\R_{2k',1}^{1,0}](q)\!\!&=&\!\!
  \big(\chit_{2k-1,1}(q)+\chit_{2k+1,1}(q)\big)\otimes\big(\chit_{2k'-1,1}(q)+\chit_{2k'+1,1}(q)\big)\nn
 \!\!&=&\!\!\sum_{j=|2k-2k'|+1,\ \!{\rm by}\ \!2}^{2(k+k')-3}\chit_{j,1}(q)
  +\sum_{j=|2k-2k'-2|+1,\ \!{\rm by}\ \!2}^{2(k+k')-1}\chit_{j,1}(q)\nn
 \!\!&+&\!\!\sum_{j=|2k-2k'+2|+1,\ \!{\rm by}\ \!2}^{2(k+k')-1}\chit_{j,1}(q)
  +\sum_{j=|2k-2k'|+1,\ \!{\rm by}\ \!2}^{2(k+k')+1}\chit_{j,1}(q)\nn
 \!\!&=&\!\!\sum_{j=|k-k'|}^{k+k'}\dkk\big(\chit_{2j-1,1}(q)+\chit_{2j+1,1}(q)\big)
\eea
while
\be
 \chit[\R_{2k,1}^{1,0}\otimes\R_{2k',1}^{1,0}](q)\ =\ 
   \sum_{j=|k-k'|}^{k+k'}\dkk\chit[\R_{2j,1}^{1,0}](q)
   \ =\  \sum_{j=|k-k'|}^{k+k'}\dkk\big(\chit_{2j-1,1}(q)+\chit_{2j+1,1}(q)\big)
\ee
The separation into a
horizontal and a vertical part (\ref{r11s}) and (\ref{kk}) then implies that 
the characters of the fundamental fusion algebra exhibit two independent 
$s\ell(2)$ structures as in (\ref{sl2}) -- one in each direction. This is clearly
reminiscent of the fusion algebras of rational (minimal) models where the
$s\ell(2)$ structures are carried by the (characters of the) {\em irreducible} representations.
Here, on the other hand, the $s\ell(2)$ structures are tied to the 
{\em Kac} representations but, due to the higher-rank indecomposable nature of some 
other representations, only at the level of their {\em characters}.

\subsection{Comparison with Eberle and Flohr}
\label{sectionEF}

To facilitate a comparison with \cite{EberleF06} by Eberle and Flohr, we  
provide a partial dictionary relating
our notation to the one used in \cite{EberleF06}. In the orders specified, the translation reads
\bea
 \{(2k,b),(1,3k)\}
   &\longleftrightarrow&\{{\cal V}(\D_{2k,b}),{\cal V}(\D_{1,3k})\},\hspace{2cm} b=1,2,3;\ k\in\mathbb{N}\nn
 \{(1,1),(1,2)\}
   &\longleftrightarrow&\{\R^{(1)}(0)_{2},\R^{(1)}(0)_{1}\}\nn
 \{\R_{2,1}^{1,0},\R_{4,1}^{1,0},\R_{6,1}^{1,0},\R_{8,1}^{1,0}\}
  &\longleftrightarrow&\{\R^{(2)}(0,2)_{7},\R^{(2)}(2,7),\R^{(2)}(7,15),\R^{(2)}(15,26)\}\nn
 \{\R_{2,2}^{1,0},\R_{4,2}^{1,0},\R_{6,2}^{1,0},\R_{8,2}^{1,0}\}
  &\longleftrightarrow&\{\R^{(2)}(0,1)_{5},\R^{(2)}(1,5),\R^{(2)}(5,12),\R^{(2)}(12,22)\}\nn
 \{\R_{2,3}^{1,0},\R_{4,3}^{1,0},\R_{6,3}^{1,0},\R_{8,3}^{1,0}\}
  &\longleftrightarrow&\{\R^{(2)}(1/3,1/3),\R^{(2)}(1/3,10/3),\R^{(2)}(10/3,28/3),\R^{(2)}(28/3,55/3)\}\nn
 \{\R_{1,3}^{0,1},\R_{1,6}^{0,1},\R_{1,9}^{0,1},\R_{1,12}^{0,1}\}
  &\longleftrightarrow&\{\R^{(2)}(0,1)_{7},\R^{(2)}(2,5),\R^{(2)}(7,12),\R^{(2)}(15,22)\}\nn
 \{\R_{2,3}^{0,1},\R_{2,6}^{0,1},\R_{2,9}^{0,1},\R_{2,12}^{0,1}\}
  &\longleftrightarrow&\{\R^{(2)}(1/8,1/8),\R^{(2)}(5/8,21/8),\R^{(2)}(33/8,65/8),\R^{(2)}(85/8,133/8)\}\nn
 \{\R_{1,3}^{0,2},\R_{1,6}^{0,2},\R_{1,9}^{0,2},\R_{1,12}^{0,2}\}
  &\longleftrightarrow&\{\R^{(2)}(0,2)_{5},\R^{(2)}(1,7),\R^{(2)}(5,15),\R^{(2)}(12,26)\}\nn
 \{\R_{2,3}^{0,2},\R_{2,6}^{0,2},\R_{2,9}^{0,2},\R_{2,12}^{0,2}\}
  &\longleftrightarrow&\{\R^{(2)}(5/8,5/8),\R^{(2)}(1/8,33/8),\R^{(2)}(21/8,85/8),\R^{(2)}(65/8,161/8)\}\nn
 \{\R_{2,3}^{1,1},\R_{4,3}^{1,1},\R_{6,3}^{1,1},\R_{8,3}^{1,1}\}
  &\longleftrightarrow&\{\R^{(3)}(0,0,1,1),\R^{(3)}(0,1,2,5),\R^{(3)}(2,5,7,12),\R^{(3)}(7,12,15,22)\}\nn
 \{\R_{2,3}^{1,2},\R_{4,3}^{1,2},\R_{6,3}^{1,2},\R_{8,3}^{1,2}\}
  &\longleftrightarrow&\{\R^{(3)}(0,0,2,2),\R^{(3)}(0,1,2,7),\R^{(3)}(1,5,7,15),\R^{(3)}(5,12,15,26)\}\nn
\label{dictEF}
\eea
The only three fusions of rank-3 indecomposable representations considered in
\cite{EberleF06} correspond to
\bea
 \R_{2,3}^{1,1}\otimes \R_{2,3}^{1,1}\!\!&=&\!\! \R_{2,3}^{1,1}\oplus2\R_{4,3}^{1,1}\oplus
  \R_{6,3}^{1,1}\oplus4\R_{2,3}^{1,2}\oplus2\R_{4,3}^{1,2}
   \oplus8\R_{2,3}^{1,0}\oplus4\R_{4,3}^{1,0} \nn
 \R_{2,3}^{1,1}\otimes \R_{2,3}^{1,2}\!\!&=&\!\! 4\R_{2,3}^{1,1}\oplus2\R_{4,3}^{1,1}\oplus
   \R_{2,3}^{1,2}\oplus2\R_{4,3}^{1,2}\oplus\R_{6,3}^{1,2}\oplus2\R_{2,3}^{1,0}\oplus
   4\R_{4,3}^{1,0}\oplus2\R_{6,3}^{1,0}  \nn
 \R_{2,3}^{1,2}\otimes \R_{2,3}^{1,2}\!\!&=&\!\! \R_{2,3}^{1,1}\oplus2\R_{4,3}^{1,1}\oplus
   \R_{6,3}^{1,1}\oplus4\R_{2,3}^{1,2}\oplus2\R_{4,3}^{1,2}\oplus2\R_{2,3}^{1,0}\oplus
   2\R_{4,3}^{1,0}\oplus2\R_{6,3}^{1,0}\oplus\R_{8,3}^{1,0} 
\label{REF}
\eea
Likewise, the only fusion of the type (\ref{ex22}) considered in \cite{EberleF06} corresponds to
\be
  \R_{2,2}^{1,0}\otimes \R_{2,3}^{0,2}\ =\ 2\R_{2,3}^{0,1}\oplus\R_{4,3}^{0,1}
   \oplus(2,3)\oplus2(4,3)\oplus(6,3)
\ee
We find that our fusion rules reduce to the many examples examined by Eberle and Flohr \cite{EberleF06}.
This confirms their observation that indecomposable representations of rank 3
are required. Our results also demonstrate that the fusion algebra closes
without the introduction of indecomposable representations of higher rank
than 3.

Eberle and Flohr also presented an algorithm \cite{EberleF06} for computing fusion products in
the augmented $c_{p,p'}$ models, in particular in the augmented $c_{2,3}=0$ model.
Their algorithm is rooted in the many explicit examples examined in their paper and yields
fusion rules which are both commutative and associative.
Considering the affirmative comparison of our fusion rules with their examples, we believe
that their algorithm for the augmented $c_{2,3}$ model
yields results equivalent to our explicit fusion rules for critical percolation ${\cal LM}(2,3)$.

\subsection{Kac Representations Revisited}

As already indicated and also discussed in \cite{EberleF06}, 
the two representations $(1,1)$ and $(1,2)$ (there denoted
$\R^{(1)}(0)_{2}$ and $\R^{(1)}(0)_{1}$, respectively) are not fully reducible. We quote Eberle and Flohr:
\begin{quote}
On the other hand, the representations $\R^{(2)}(0,1)_5$ and $\R^{(2)}(0,1)_7$
contain a state with weight 0 which generates a subrepresentation $\R^{(1)}(0)_1$.
This subrepresentation is indecomposable but neither is it irreducible nor does it exhibit
any higher rank behaviour. It only exists as a subrepresentation as it needs the embedding 
into the rank 2 representation in order not to have nullvectors at both levels 1 and 2.
But, nevertheless, being a subrepresentation of a representation in the spectrum 
it has to be included into the spectrum, too.
\end{quote}
This is corroborated by our findings. {}From the lattice, the two representations
$(1,1)$ and $(1,2)$ arise in the conformal
scaling limit from very simple and natural boundary conditions. 
This supports our assertion that these Kac representations are indeed
physical. Furthermore, since one is immediately faced with problems when attempting to include
their irreducible components 
\be
 (1,1):\ \ \{{\cal V}(0),{\cal V}(2)\},\hspace{2cm}
 (1,2):\ \ \{{\cal V}(0),{\cal V}(1)\}
\ee
in the fusion algebra, we advocate to consider fusion algebras of critical percolation
generated from Kac representations and indecomposable
representations of higher rank. The only irreducible representations appearing in these
fusion algebras are therefore themselves Kac representations, that is, they belong to the set of
irreducible Kac representations $\{(2k,1),(2k,2),(2k,3)=(2,3k),(1,3k)\}$.
Natural extensions of the horizontal, vertical and fundamental
fusion algebras involve {\em all} the associated Kac representations and read
\be
 \big\langle(2,1),(3,1)\big\rangle,
 \hspace{1cm}\big\langle(1,2),(1,4)\big\rangle,
 \hspace{1cm}\big\langle(2,1),(3,1),(1,2),(1,4)\big\rangle
\label{full}
\ee
respectively. They will be addressed elsewhere. Further evidence in support of
the relevance of Kac representations in logarithmic CFT may be found in \cite{Ras}
where quotient modules with characters (\ref{chikac}) are found to arise naturally in
the limit of certain sequences of minimal models.

\section{Conclusion}

We have presented explicit general conjectures for the chiral fusion algebras of critical percolation,
and we have exhibited  dictionaries to facilitate comparison of our results with the particular results of Eberle and Flohr~\cite{EberleF06} and Read and Saleur~\cite{RS}. 
Importantly, we observe the appearance of rank-3 indecomposable representations in agreement with Eberle and Flohr. 
Our fundamental fusion algebra is built from independent horizontal and vertical algebras that, 
at the level of characters, respect an underlying $s\ell(2)$ structure. 
The identity $(1,1)$ of this fundamental fusion algebra
is a reducible yet indecomposable Kac representation of rank 1.

Our reported fusion rules are supported by extensive numerical investigations of an integrable lattice model of critical percolation. 
These lattice results will be presented elsewhere. We also hope to discuss elsewhere
the full fusion algebra encompassing all of the Kac representations as well as extensions to general logarithmic minimal models.
\vskip.5cm
\section*{Acknowledgments}
\vskip.1cm
\noindent
This work is supported by the Australian Research Council. 
JR thanks Andreas Ludwig for encouraging discussions at the KITP in November 2006.



\begin{thebibliography}{99}

\bib{Kesten82} H. Kesten, {\em Percolation Theory for Mathematicians}, Birkh\"auser, Boston, 1982.

\bib{Grimmet89} G. Grimmet, {\em Percolation}, Springer-Verlag, New York, 1989.

\bib{Stauffer92} D. Stauffer and A. Aharony, {\em Introduction to Percolation Theory}, Taylor and Francis, London, 1992.

\bib{BroadHamm57} S. Broadbent and J. Hammersley, 
 Proc. Camb. Phil. Soc. {\bf 53} (1957) 629-641.

\bib{Cardy92} J.L. Cardy,
J. Phys. {\bf A25} (1992) L201--L206.

\bib{Smirnov01} S. Smirnov,
C. R. Acad. Sci. Paris {\bf 333} (2001) 239--244.

\bib{Schramm00} O. Schramm,
Israel J. Math. {\bf 118} (2000) 221--288.

\bib{LawlerEtAl01} G.F. Lawler, O. Schramm and W. Werner,
Acta Math. {\bf 187} (2001) 237--273.

\bib{Werner03} W. Werner,
 {\em Random planar curves and Schramm-Loewner evolutions},
arXiv:math/0303354.

\bib{RohdeEtAl05} S. Rohde and O. Schramm,
Ann. Math. {\bf 161} (2005) 879--920.

\bib{KN} W. Kager and B. Nienhuis, J. Stat. Phys. {\bf 115} (2004) 1149-1229.

\bib{BB} M. Bauer and D. Bernard, Phys. Rept. {\bf 432} (2006) 115-221.

\bib{RazStrog01} A.V. Razumov and Yu. G. Stroganov,
J. Phys. {\bf A34} (2001) 5335--5340.

\bib{BatchelorEtAl01} M.T. Batchelor, J. de Gier and B. Nienhuis,
J. Phys. {\bf A34} (2001) L265-L270.

\bib{PearceEtAl02} P.A. Pearce, V. Rittenberg, J. de Gier and B. Nienhuis,
J. Phys. {\bf A35} (2002) L661--L668.

\bib{FZZ06} P. Di Francesco, P. Zinn-Justin and J.-B. Zuber, J. Stat. Mech. (2006) P08011.

\bib{Cardy99} J. Cardy, {\em Logarithmic correlations in quenched random magnets and polymers},
arXiv:cond-mat/9911024.

\bib{GL} V. Gurarie and A.W.W. Ludwig, J. Phys. {\bf A35} (2002) L377-L384;
{\em Conformal field theory at central charge $c=0$
 and two-dimensional critical systems with quenched disorder}, arXiv:hep-th/0409105.

\bib{FL} M. Flohr and A. M\"uller-Lohmann, J. Stat. Mech. (2006) P04002.

\bibitem{Roh96} F.~Rohsiepe, {\it On reducible but indecomposable representations of the
Virasoro algebra}, arXiv:hep-th/9611160.

\bibitem{Gurarie93} V.~Gurarie, 
Nucl. Phys. {\bf B410} (1993) 535-549.


\bib{MS} Z. Maassarani and D. Serban, Nucl. Phys. {\bf B489} (1997) 603-625.

\bibitem{Flohr97}
M.~Flohr, 
Int. J. Mod. Phys. {\bf A12} (1997) 1943--1958.

\bib{RAK} M.R. Rahimi Tabar, A. Aghamohammadi and M. Khorrami, 
 Nucl. Phys. {\bf B497} (1997) 555-566.

\bib{Kausch00} H.G. Kausch, Nucl. Phys. {\bf B583} (2000) 513-541.

\bibitem{bdylcft}  I.I.~Kogan and J.F.~Wheater, 
Phys. Lett. {\bf B486} (2000) 353-361.

\bib{MRS} S. Moghimi-Araghi, S. Rouhani and M. Saadat, Nucl. Phys. {\bf B599} (2001) 531-546.

\bib{Flohr03} 
M. Flohr, Int. J. Mod. Phys. {\bf A18} (2003) 4497--4592.

\bib{Gaberdiel03} 
M.R. Gaberdiel, Int. J. Mod. Phys. {\bf A18} (2003) 4593--4638.

\bib{FFHST} J. Fjelstad, J. Fuchs, S. Hwang, A.M. Semikhatov and I.Yu. Tipunin,
 Nucl. Phys. {\bf B633} (2002) 379-413.

\bib{Ruelle02} P. Ruelle, Phys. Lett. {\bf B539} (2002) 172-177.

\bib{Kawai03}
S. Kawai, Int. J. Mod. Phys. {\bf A18} (2003) 4655--4684.

\bib{LMRS} F. Lesage, P. Mathieu, J. Rasmussen and H. Saleur, Nucl. Phys. {\bf B647} (2002) 363-403;
 Nucl. Phys. {\bf B686} (2004) 313-346.

\bib{Nichols02}
A. Nichols, {\em $SU(2)_k$ logarithmic conformal field theories}, Ph.D. thesis, University of Oxford, 
 arXiv:hep-th/0210070.

\bib{RasWZW} J. Rasmussen, Nucl. Phys. {\bf B736} (2006) 225-258.

\bib{FG} M. Flohr and M.R. Gaberdiel, J. Phys. {\bf A39} (2006) 1955-1968.

\bib{FGST} B.L. Feigin, A.M. Gainutdinov, A.M. Semikhatov and I.Yu. Tipunin, 
 Nucl. Phys. {\bf B757} (2006) 303-343.

\bib{GR} M.R. Gaberdiel and I. Runkel, J. Phys. {\bf A39} (2006) 14745-14780.

\bib{QS} T. Quella and V. Schomerus, {\em Free fermion resolution of supergroup  WZNW models},
 arXiv:0706.0744.



\bib{SaleurD87} H. Saleur and B. Duplantier,
Phys. Rev. Lett. {\bf 58} (1987) 2325--2328.

\bib{SaleurSUSY} H.~Saleur, Nucl. Phys. {\bf B382} (1992) 486--531.

\bib{ReSa01} N. Read and H. Saleur, Nucl. Phys. {\bf B613} (2001) 409-444. 

\bib{PRZ} P.A.~Pearce, J.~Rasmussen and J.-B.~Zuber, 
J. Stat. Mech. (2006) P11017.

\bib{PRpoly} P.A. Pearce and J. Rasmussen, 
J. Stat. Mech. (2007) P02015.

\bib{EberleF06} H. Eberle and M. Flohr, J. Phys. {\bf A39} (2006) 15245-15286. 

\bib{GK} M.R.~Gaberdiel and H.G.~Kausch, 
Nucl. Phys. {\bf B477} (1996) 293--318.

\bib{Nahm94} W.~Nahm, 
Int. J. Mod. Phys. {\bf B8} (1994) 3693-3702.

\bib{RS} N. Read and H. Saleur, {\em Associative-algebraic approach to logarithmic conformal field
 theories}, arXiv:hep-th/0701117.
 

\bib{FSZ} P. Di Francesco, H. Saleur and J.-B. Zuber, Nucl. Phys. {\bf B285} (1987) 454-480.

\bib{Ras} J. Rasmussen, Nucl. Phys. {\bf B701} (2004) 516-528; Int. J. Mod. Phys. {\bf A22} (2007) 67-82.


\end{thebibliography}
\end{document}